\documentclass[12pt]{article}
\usepackage{amsmath, amssymb, slashed}
\usepackage{xcolor}
\usepackage[colorlinks=true,urlcolor=black,linkcolor=black,citecolor=black]{hyperref}
\usepackage[utf8]{inputenc}

\newcommand{\R}{\mathbb{R}}
\newcommand{\Ldelta}{{\delta_{\mathsf{l}}}}
\newcommand{\Rdelta}{{\delta_{\mathsf{r}}}}
\newcommand{\Ur}{\mathsf{U}_{\mathsf{r}}}
\newcommand{\Ul}{\mathsf{U}_{\mathsf{l}}}
\newcommand{\wk}[1]{\mathord{:} #1 \mathord{:}}

\newtheorem{lem}{Lemma}

\begin{document}
\title{Stochastic quantization of massive fermions}

\author{
A.N. Efremov
\\
CPHT, Ecole Polytechnique, CNRS, Université Paris-Saclay,\\
Route de Saclay, 91128 Palaiseau, France.
\\
\href{mailto:alexander@efremov.fr}{alexander@efremov.fr}
}

\maketitle

\begin{abstract}We consider a general solution of the Langevin equation describing massive fermions to an appropriate boundary problem. Assuming existence of such solution we show that its correlators coincide with the Schwinger functions of corresponding Euclidean Quantum Field Theory.
\end{abstract}
\section{Introduction} In this work we make an effort to fill a gap in literature and to establish a mathematically sound connection between the Langevin equation and massive fermionic models in the quantum field theory. A motivation for this analysis is our desire to extend the methods developed recently by A.Kupiainen~\cite{kup} and M.Hairer~\cite{hai} to non-abelian gauge theories, Yukawa and Gross--Neveu models~\cite{Glimm1967,gross_neveu}. Compared with a functional integral approach stochastic PDE's present a clear simplification and allow us to consider a much wider class of models. We hope that the problem of \textit{criticality} which puts severe limitation on the dimension of time-space can be solved using Wilson's non-perturbative renormalization group~\cite{wilson}. A common way to perform stochastic quantization is of course the Fokker--Planck equation, see Parisi and Wu~\cite{parwu}. Unfortunately a fermionic field is a function in infinite-dimensional Grassmann algebra~\cite{ber} and despite its random or stochastic character, is not a stochastic process as we understand it in the probability theory. Although the Fokker--Planck approach suggested in~\cite{damgaard} for fermions is simple, intuitive and better suited for an initial reading we should keep in mind that the whole description remains at a formal level only. On the other hand the functional integral approach makes our construction for fermions mathematically meaningful. Rather than formally derive the Fokker--Planck equation for fermions we follow the functional integral approach proposed by J.~Zinn-Justin for the scalar~$\phi^4$ model~\cite{zinn86}.

The first important object is the Euclidean action. Without any loss of generality we consider Dirac spinors in two dimensions. The 2-points Schwinger function for Dirac spinors can be obtained from the corresponding Wightman function on an appropriately chosen subspace by the Wick rotation~\cite{Osterwalder1973}, i.e. by the mapping $x_1 \mapsto -i x_1$,
\begin{align}
W(x)&=\int \frac{d^2 p}{(2 \pi)^2} \frac{ -i \hat{\slashed{p}} +m}{p^2 + m^2 - i \epsilon} e^{i p x},\\
S(x)&=\int \frac{d^2 p}{(2 \pi)^2} \frac{ -i \slashed{p} +m}{p^2 + m^2} e^{i p x}.\label{2212a}
\end{align}
Here $\hat{\slashed{p}}=-p_1  \hat{\gamma}_1 + p_2  \gamma_2$, $\hat{\gamma_1}=-i \gamma_1$, $\slashed{p}=p_k  \gamma_k$, $\{\gamma_i, \gamma_j\}=2 \delta_{ij}$, $\gamma^+_i=\gamma_i$. The matrices $\hat{\gamma}_i$, $\gamma_i$ correspond to Minkowski and Euclidean space-time respectively. We choose the representation for the matrices~$\gamma_i$ such that they coincide with the Pauli matrices:
\begin{align}
\gamma_1&=\begin{pmatrix}0,&1\\1,&0\end{pmatrix}&\gamma_2&=\begin{pmatrix}0,&-i\\i,&0\end{pmatrix}&\gamma_3&=\begin{pmatrix}1,&0\\0,&-1\end{pmatrix}
\end{align}
Hence~$\gamma_1^T=\gamma_1$, $\gamma_3=-i\gamma_1 \gamma_2$. It is easy to see that the function in~\eqref{2212a} is a fundamental solution to a non-homogeneous Dirac equation in Euclidean space-time, i.e.
\begin{equation}
(\slashed{\partial} + m)S(x)=\delta(x). 
\end{equation}
As an example we consider the Yukawa model~\cite{Glimm1967}. The model involves a scalar field~$\phi$ and a fermion field~$\psi$,
\begin{equation}
  L = \int \limits d^2 x \; \bar{\psi}(\slashed{\partial}+m)\psi + g \wk{\bar{\psi}\psi} \phi  - \frac{1}{2} \phi \Delta \phi + \frac{1}{2} M^2 \phi^2  ,\label{0902a}
\end{equation}
where $\psi, \bar{\psi}: \R^2 \to \Lambda^2$ and $\phi: \R^2 \to \R$ are respectively independent spinor and scalar fields. Here by $\Lambda$ we denote an infinite-dimensional Grassmann algebra. Since the expectation $\langle \bar{\psi} \psi \rangle$ turns out to be divergent we should introduce a regularization to properly define the Wick product~$\wk{\bar{\psi}\psi}$. By definition the Wick product includes the ordinary product $\bar{\psi}\psi$ and a counterterm~$c$ which becomes singular when we remove the regularization, i.e. $\wk{\bar{\psi}\psi}=\bar{\psi}\psi - c$. The matrices $\gamma_i$ transform as vectors under the rotation $u \gamma_i u^{-1}$ where $u=e^{i \gamma_3\frac{\alpha}{2}}$. Since $\gamma_3$ is hermitian $u$ is unitary. The quantities $\bar{\psi} \psi$ and $\bar{\psi}  \slashed{\partial} \psi$ are scalars, i.e. invariant with respect to the action $\bar{\psi} \to \bar{\psi}u^+$ and $\psi \to u \psi$, in appendix~\ref{0902b} we state other properties of the action.

The second important object is the Langevin equation:
\begin{align}
\partial_t \phi &=-\frac{\delta L}{\delta \phi} + \xi =  \Delta \phi - M^2 \phi - g \wk{\bar{\psi} \psi} + \xi ,\label{2101a}\\
\partial_t \psi &=-\frac{\delta L}{\delta \bar{\psi}} + \eta =  -(\slashed{\partial}+m)\psi - g \psi \phi + \eta,\\
\partial_t \bar{\psi} &=\frac{\delta L}{\delta \psi} + \bar{\eta} =  ( \slashed{\partial}^T-m)\bar{\psi}  - g \bar{\psi} \phi + \bar{\eta} \label{2101b},
\end{align}
where $\xi$, $\eta$, $\bar{\eta}$ denote the corresponding Gaussian white noise. Having a solution of these equations one can calculate different correlators. Below we show that such correlators coincide with the Schwinger functions which we obtain using the corresponding generating functional. 
\section{Stochastic quantization of massive fermions}
First let us write the fundamental solution of the linear equations corresponding to the retarded Green functions
\begin{align}
(\partial_t + \slashed{\partial} + m)G&=\delta(t,x),\\
(\partial_t - \slashed{\partial}^T + m)\bar{G}&=\delta(t,x),\\
(\partial_t - \Delta + M^2) P&=\delta(t,x).
\end{align}
These functions vanish $\forall t<0$, i.e. $P(t,x)=0$, $G(t,x)=0$ and $\bar{G}(t,x)=0$. 
\begin{align}
G(t,x)&=\frac{1}{i}\int \frac{d w d^2 p}{(2\pi)^3}  \frac{w -im - \slashed{p}}{(w-im)^2 - p^2} e^{iwt +ipx},\\
\bar{G}(t,x)&=\frac{1}{i}\int \frac{d w d^2 p}{(2\pi)^3}  \frac{w -im + \slashed{p}^T}{(w-im)^2 - p^2} e^{iwt +ipx},\\
P(t,x)&=\frac{1}{i}\int \frac{d w d^2 p}{(2\pi)^3}  \frac{1}{w - i (p^2 + M^2)} e^{iwt +ipx}.
\end{align}
Here $\bar{G}=\gamma_2 G \gamma_2$. 
We can write equations \eqref{2101a}-\eqref{2101b} in the following form 
\begin{equation}
\partial_t \Psi + \tilde{\mathsf{D}} \Psi + \mathsf{V}(\Psi) - \Xi=0\,,\label{405a}
\end{equation}
where
\begin{align}
\Xi&=\begin{pmatrix}\eta\\\bar{\eta}\\\xi\end{pmatrix},&\Psi&= \begin{pmatrix}\psi\\\bar{\psi}\\\phi\end{pmatrix},&\tilde{\mathsf{D}}&=\begin{pmatrix}\slashed{\partial}+m\\-\slashed{\partial}^T+m\\-\Delta + M^2\end{pmatrix}, &\mathsf{V}(\Psi)&= g\begin{pmatrix}\psi \phi \\ \bar{\psi}\phi\\ \wk{\bar{\psi}\psi} \end{pmatrix}\,.
\end{align}
Here $\Psi(t,x)$ is a distribution in $(a,b) \times \mathbb{R}^2$ which satisfies an initial value problem~$\Psi(t_0,x)=\Psi_0(x)$ for equation~\eqref{405a}, $t_0 \in (a,b) \subset \mathbb{R}$. Then we put equation~\eqref{405a} into an integral form
\begin{align}
\Psi&= \mathsf{G}_{[+\infty ,t_0]}(-\mathsf{V} + \Xi)+ e^{-\tilde{\mathsf{D}}(t-t_0)}\Psi_0,&\mathsf{G}&=\begin{pmatrix}G&0&0\\0&\bar{G}&0\\0&0&P\end{pmatrix} \label{1805a}
\end{align}
\begin{equation}
(\mathsf{G}_{[+\infty ,t_0]}f)(t,x)=\int \limits^{\infty}_{t_0} ds \int  d^2 z \, \mathsf{G}(t-s,x-z) f(s, z)\,.  
\end{equation}
This result follows immediately from the definition of the retarded Green function~$\mathsf{G}$, i.e. $(\partial_t + \tilde{\mathsf{D}})\mathsf{G}=\delta(t,x)$. We also assume that a solution for the initial value problem exists for an infinite time interval, $t_0=-\infty$. Consequently we can identify the time interval~$(a,t_0)$ with $\mathbb{R}$. When $t_0=-\infty$, since $m>0$ it follows from~\eqref{1805a} that $\Psi(t,x)$ should solve a fix point problem
\begin{align}
\Psi&= \mathsf{G}(-\mathsf{V} + \Xi)\,. \label{2101c}
\end{align}
Here and everywhere in the text below $\mathsf{G}f=\mathsf{G}_{[+\infty,-\infty]}f$. Under the above assumption and an appropriate regularity of the potential~$\mathsf{V}$ the choice of model, the Yukawa model in two dimensions in our case, is not important in the coming lemmas whose proofs hold in general, so we try to keep general notations as much as possible.

Following the great success in the construction of the massive scalar model using methods of functional integral~\cite{gljf} K.Osterwalder and R.Schrader constructed the Euclidean theory of free fermions~\cite{PhysRevLett.29.1423}. To put it simply, given an action~$L$ which involves fermions one defines a generating functional~$Z_{ft}$ for connected Schwinger functions using an integral over infinite dimensional Grassmann algebra~\cite{ber} along with the usual functional integral for bosons. Denoting these fields by $\tilde{\Psi}=(\tilde{\psi}, \tilde{\bar{\psi}},\tilde{\phi})$ we have
\begin{align}
Z_{ft}(\tilde{K})&= \int D \tilde{\Psi}\, e^{-L + \tilde{\Psi} Q \tilde{K}}\,,&Q&=\begin{pmatrix}0&-1&0\\1&0&0\\0&0&1\end{pmatrix},&\tilde{K}&=\begin{pmatrix}\tilde{k}\\\tilde{\bar{k}}\\\tilde{j}\end{pmatrix}\,.\label{405b}
\end{align}
Here $\tilde{k},\tilde{\psi}:\mathbb{R}^2 \to \Lambda^2$ and $\tilde{j},\tilde{\phi}:\mathbb{R}^2 \to \mathbb{R}$ in two dimensions. To shorten the notation we use $\tilde{\Psi} Q \tilde{K}$ instead of $\int d^2x \, \tilde{\Psi}(x) Q  \tilde{K}(x)$. The composite field~$\tilde{\Psi}$ does not depend on the stochastic time. Given a fixed time~$T$ we could define $\tilde{\Psi}(x)=\Psi(T,x)$ but it is not necessary here. We use a tilde to emphasize that a variable or an operator is explicitly independent of the stochastic time. For convenience we introduce the notation
\begin{equation}
\tilde{\mathsf{E}}(\Psi)=\tilde{\mathsf{D}} \Psi + \mathsf{V}(\Psi).
\end{equation}
Since the elements of Grassmann algebra are anti-commuting we will distinguish the left~$\Ldelta$ and right derivatives~$\Rdelta$.
\begin{lem} The partition function~$Z_{ft}$ satisfies the following equation
\begin{equation}
\left(\tilde{\mathsf{E}}(Q^T \frac{\delta}{\Ldelta {\tilde{K}}})-\tilde{K}\right) Z_{ft}(\tilde{K})=0.\label{0902c}
\end{equation}
\end{lem}
\emph{Proof}
\begin{align}
\tilde{K} Z_{ft}(\tilde{K})&= \int D \Psi\, e^{-L}\Big( Q^T \frac{\delta e^{\Psi Q \tilde{K}}}{\Ldelta \Psi}\Big)=\int D \Psi\, \Big(- Q^T \frac{\delta e^{-L}}{\Ldelta \Psi}\Big) e^{\Psi Q \tilde{K}}\\
&=Q^T \frac{\delta L}{\Ldelta \Psi}\Big|_{\Psi= Q^T \Ldelta_{\tilde{K}}} Z_{ft}(\tilde{K})=\tilde{\mathsf{E}}(Q^T \Ldelta_{\tilde{K}})\, Z_{ft}(\tilde{K})\,.
\end{align}
\hfill$\blacksquare$\\
Let $\Psi$ be a solution of the fix point problem given in~\eqref{2101c}. Define the following a generating functional~$Z$:
\begin{align}
Z(K)&= \int D \Xi \, e^{-\frac{1}{2}\Xi \mathsf{Q} \Xi + K Q \Psi}\,,&K&=\begin{pmatrix}k\\\bar{k}\\j\end{pmatrix}\,.\label{0902d}
\end{align}
Here $k,\psi:\mathbb{R}^{2+1} \to \Lambda^2$ and $j,\phi:\mathbb{R}^{2+1} \to \mathbb{R}$ in two dimensions. We want to show that at some finite time $t=T \in \mathbb{R}$ the generating functional~$Z(K)$ satisfies equation~\eqref{0902c}. Since $Z(0)=Z_{ft}(0)=1$ it will imply that $Z(K)=Z_{ft}(\tilde{K})$, i.e. the correlators obtained from~\eqref{0902d} at time~$T$, i.e. $\langle \Psi(T,x_1)... \Psi(T,x_n) \rangle$, coincide with the Schwinger functions of~\eqref{405b}.
\begin{lem} Let~$\Psi$ be a solution of the fix point problem stated in~\eqref{2101c}. Furthermore, let $\tilde{Z}(\tilde{K})=Z(K)$ where $K(t,x)=\delta(t-T)\tilde{K}(x)$. The partition function~$\tilde{Z}$ satisfies the equation
\begin{equation}
\left(\tilde{\mathsf{E}}(Q^T \frac{\delta}{\Ldelta \tilde{K}})-\tilde{K}\right) \tilde{Z}(\tilde{K})=0.\label{0902f}
\end{equation}
\end{lem}
\emph{Proof} For the expectation of the noise using~\eqref{0902d} we obtain
\begin{align}
\langle \Xi \rangle_K &=Q^T \frac{\delta \Psi^T}{\Ldelta \Xi}  Q K Z(K),&\frac{\delta \Psi^T}{\Ldelta \Xi}&= \mathsf{G}^T \Big(1 +  \frac{\delta \mathsf{V}^T}{\Ldelta \Psi} \mathsf{G}^T\Big)^{-1}  .\label{2501b}
\end{align}
Furthermore
\begin{align}
\partial_t \Psi=\Big(1 + \mathsf{G} \frac{\delta \mathsf{V}}{\Rdelta \Psi}\Big)^{-1} \partial_t \mathsf{G} \Xi\,.
\end{align}
Since~$\Psi$ satisfies the Langevin equation, i.e. $\partial_t \Psi +\tilde{\mathsf{E}}(\Psi) - \Xi=0$ we have
\begin{equation}
\Big[\Big(1 + \mathsf{G} \frac{\delta \mathsf{V}}{\Rdelta \Psi}\Big)^{-1} \partial_t \mathsf{G} -1  \Big] \Xi + \tilde{\mathsf{E}}(\Psi)=0\,.\label{2501a}
\end{equation}
Calculating the expectation of \eqref{2501a} and using \eqref{2501b} we find
\begin{equation}
\Big(\tilde{\mathsf{E}} + \Big[\Big(1 + \mathsf{G} \frac{\delta \mathsf{V}}{\Rdelta \Psi}\Big)^{-1} \partial_t \mathsf{G} -1 \Big]Q^T \mathsf{G}^T \Big(1 +  \frac{\delta \mathsf{V}^T}{\Ldelta \Psi}\mathsf{G}^T\Big)^{-1} Q K \Big)Z(K)=0\,.\label{2501d}
\end{equation}
We restrict our interest to the correlators at $t=T$, i.e. $K=\delta(t - T)\tilde{K}(x)$. Since $\mathsf{G}(t,x)=0$ for $t<0$ we have
\begin{equation}
\lim_{t \to 0}\Big[\mathsf{G}^T\Big(1 + \frac{\delta \mathsf{V}^T}{\Ldelta \Psi} \mathsf{G}^T\Big)^{-1}\Big]_{t,x}=\lim_{t \to 0}\mathsf{G}^T(t,x)=\delta(x).
\end{equation}
Defining
\begin{align}
\Ur&=\frac{\delta \mathsf{V}}{\Rdelta \Psi},&\tilde{R}&=\Big(1 + \mathsf{G} \Ur  \Big)^{-1} \partial_t \mathsf{G} Q^T \mathsf{G}^T \Big(1 +  \Ul^T \mathsf{G}^T\Big)^{-1}\,,\label{2601a}
\end{align}
we rewrite equation~\eqref{2501d} in the form 
\begin{equation}
(\tilde{\mathsf{E}} - \tilde{K} + \tilde{R} Q \tilde{K})Z(K)=0 \label{2501f}
\end{equation}
It remains to prove that equations~\eqref{2501f} and~\eqref{0902f} are equivalent. One can show that for the Green functions the following holds
\begin{equation}
\partial_t \mathsf{G}_{t_1-\tau,x-z} Q^T \mathsf{G}^T_{\tau-t_2, z-y}=\frac{Q^T\mathsf{G}^T_{t_1 - t_2,x-y} - \mathsf{G}_{t_1 - t_2,x-y} Q^T}{2}\,.
\end{equation}
This quantity vanishes whenever $t_1=t_2=T$. Expanding the inverse operators appearing in~\eqref{2601a} in power series over~$\mathsf{U}$ we obtain for a fixed order~$m$ the following expression
\begin{multline}
\sum \limits^{m}_{n=0}(\mathsf{G}\Ur)^n(Q^T \mathsf{G}^T - \mathsf{G}Q^T)(\Ul^T\mathsf{G}^T)^{m-n}= \sum \limits^{m}_{n=1}(\mathsf{G}\Ur)^nQ^T \mathsf{G}^T (\Ul^T\mathsf{G}^T)^{m-n}  \\-  \sum \limits^{m-1}_{n=0} (\mathsf{G}\Ur)^n \mathsf{G}Q^T(\Ul^T\mathsf{G}^T)^{m-n}  + Q^T \mathsf{G}^T (\Ul^T\mathsf{G}^T)^m - (\mathsf{G}\Ur)^m \mathsf{G}Q^T.
\end{multline}
Here the last two terms vanish if the time argument on the both ends is the same. Thus for $m>0$ this expression becomes
\begin{equation}
- \sum \limits^{m-1}_{n=0} (\mathsf{G}\Ur)^n \mathsf{G}[Q^T\Ul^T - \Ur Q^T]\mathsf{G}^T(\Ul^T\mathsf{G}^T)^{m-1-n}.
\end{equation}
After summing up all orders~$m$ we obtain the final expression for $\tilde{R}$ at equal time on the both ends
\begin{align}
\tilde{R}&=\Big(1 + \mathsf{G} \Ur  \Big)^{-1} \mathsf{G}[Q^T\Ul^T - \Ur Q^T]\mathsf{G}^T \Big(1 +  \Ul^T \mathsf{G}^T\Big)^{-1}\,.
\end{align}
The quantity $Q^T\Ul^T - \Ur Q^T$ vanishes, see appendix~\ref{2901a}. Consequently the difference between equations~\eqref{2501f} and~\eqref{0902f} vanishes.
\hfill$\blacksquare$
\section{Acknowledgements}
I thank the Institute for Theoretical Physics at the University of Leipzig, Germany for the financial support.

\appendix
\section{$\Ur$ and $\Ul^T$}\label{2901a}
Since in a general situation the potential~$\mathsf{V}$ includes all required counterterms which are fine-tuned to cancel singularities we still use here the same notation as if we have the whole action~$L$. Furthermore all derivatives below are left derivatives.
\begin{align}
\Ur&=\begin{pmatrix}-\frac{\delta^2 L}{\delta \bar{\psi}_i \delta \psi_j}&-\frac{\delta^2 L}{\delta \bar{\psi}_i \delta \bar{\psi}_j}&\frac{\delta^2 L}{\delta \bar{\psi}_i \delta \phi_j}\\
\frac{\delta^2 L}{\delta \psi_i \delta \psi_j}&\frac{\delta^2 L}{\delta \psi_i \delta \bar{\psi}_j}&-\frac{\delta^2 L}{\delta \psi_i \delta \phi_j}\\
-\frac{\delta^2 L}{\delta \phi_i \delta \psi_j}&-\frac{\delta^2 L}{\delta \phi_i \delta \bar{\psi}_j}&\frac{\delta^2 L}{\delta \phi_i \delta \phi_j}\end{pmatrix}=\begin{pmatrix} u_{11}& u_{12}& u_{13} \\ u_{21}&u_{22}&u_{23}\\u_{31}&u_{32}&u_{33}\end{pmatrix}\\
\Ul^T&=\begin{pmatrix}\frac{\delta^2 L}{\delta \psi_i \delta \bar{\psi}_j}&-\frac{\delta^2 L}{\delta \psi_i \delta \psi_j}&\frac{\delta^2 L}{\delta \psi_i \delta \phi_j}\\
\frac{\delta^2 L}{\delta \bar{\psi}_i \delta \bar{\psi}_j}&-\frac{\delta^2 L}{\delta \bar{\psi}_i \delta \psi_j}&\frac{\delta^2 L}{\delta \bar{\psi}_i \delta \phi_j}\\
\frac{\delta^2 L}{\delta \phi_i \delta \bar{\psi}_j}&-\frac{\delta^2 L}{\delta \phi_i \delta \psi_j}&\frac{\delta^2 L}{\delta \phi_i \delta \phi_j}\end{pmatrix}=\begin{pmatrix} u_{22}& -u_{21}& -u^T_{31} \\ -u_{12}&u_{11}&-u^T_{32}\\u^T_{13}&u^T_{23}&u_{33}\end{pmatrix}
\end{align}
The equation $Q^T \Ul^T=\Ur Q^T$ implies $u_{31}=u^T_{23}$ and $u_{13}=-u^T_{32}$.
\section{Symmetries of the action} \label{0902b}
We define as usual the time and parity reversal operators
\begin{align}
\mathcal{T}&:\begin{pmatrix} x_1 , x_2 \end{pmatrix} \mapsto \begin{pmatrix} -x_1 , x_2 \end{pmatrix},&\mathcal{P}&:\begin{pmatrix} x_1 , x_2 \end{pmatrix} \mapsto \begin{pmatrix} x_1 , - x_2 \end{pmatrix},
\end{align}
and corresponding transformations for the spinor~$\psi$
\begin{align}
P&: \psi(x) \mapsto \gamma_1 \psi(\mathcal{P} x),&T&: \psi(x) \mapsto \gamma_3 \psi(\mathcal{T} x),&C&: \psi(x) \mapsto \gamma_1 \bar{\psi}(x),
\end{align}
where $T$ is anti-linear, i.e. $T \alpha \psi T^{-1}=\alpha^* T \psi T^{-1}$. The action $L$, see \eqref{0902a}, is invariant under $P$. Invariance under $CT$ can be obtained if one simultaneous makes inversion of the sign of~$m$
\begin{align}
\psi(x) &\mapsto \gamma_2 \bar{\psi}(\mathcal{T}x),&\bar{\psi}(x) &\mapsto  \psi(\mathcal{T}x) \gamma_2,&m&\mapsto -m.
\end{align}
\bibliographystyle{unsrt}
\bibliography{gn}

\begin{thebibliography}{10}

\bibitem{kup}
A.~Kupiainen.
\newblock Renormalization group and stochastic pde's.
\newblock {\em Annales Henri Poincaré}, 17, 2014.

\bibitem{hai}
M.~Hairer.
\newblock Introduction to regularity structures.
\newblock {\em Braz. J. Probab. Stat.}, 29:175–210, 2015.

\bibitem{Glimm1967}
J.~Glimm.
\newblock The yukawa coupling of quantum fields in two dimensions. ii.
\newblock {\em Communications in Mathematical Physics}, 6(1):61--76, Mar 1967.

\bibitem{gross_neveu}
D.~J. Gross and A.~Neveu.
\newblock Dynamical symmetry breaking in asymptotically free field theories.
\newblock {\em Phys. Rev. D}, 10:3235--3253, Nov 1974.

\bibitem{wilson}
K.G. Wilson.
\newblock Renormalization group and critical phenomena.
\newblock {\em Phys. Rev. B}, 4(9):3174--3183, 1971.

\bibitem{ber}
F.A. Berezin.
\newblock {\em Introduction to Superanalysis}.
\newblock Springer, 1987.

\bibitem{parwu}
G.~Parisi and Y.~Wu.
\newblock Perturbation theory without gauge fixing.
\newblock {\em Scientia Sinica}, 24:483, 1981.

\bibitem{damgaard}
P.~H. Damgaard and H.~Hüffel.
\newblock Stochastic quantization.
\newblock {\em Physics Reports}, 152(5):227 -- 398, 1987.

\bibitem{zinn86}
J.~Zinn-Justin.
\newblock {Renormalization and Stochastic Quantization}.
\newblock {\em Nucl. Phys.}, B275:135--159, 1986.

\bibitem{Osterwalder1973}
K.~Osterwalder.
\newblock {\em Euclidean fermi fields}, pages 326--331.
\newblock Springer Berlin Heidelberg, Berlin, Heidelberg, 1973.

\bibitem{PhysRevLett.29.1423}
K.~Osterwalder and R.~Schrader.
\newblock Feynman-kac formula for euclidean fermi and bose fields.
\newblock {\em Phys. Rev. Lett.}, 29:1423--1425, Nov 1972.

\bibitem{gljf}
J.~Glimm and A.~Jaffe.
\newblock {\em Quantum physics}.
\newblock Springer, 1987.

\end{thebibliography}

\end{document}